\documentclass[10pt,twocolumn,letterpaper]{article}

\usepackage{iccv}              

\definecolor{iccvblue}{rgb}{0.21,0.49,0.74}
\usepackage[pagebackref,breaklinks,colorlinks,allcolors=iccvblue]{hyperref}

\usepackage{url}
\usepackage{graphicx}
\usepackage{bm}
\usepackage{amssymb}
\usepackage{multirow}
\usepackage{booktabs}
\usepackage{pifont}
\usepackage{array}
\usepackage{caption}
\usepackage{wrapfig}

\usepackage{makecell}
   \usepackage{float}

\def\paperID{303} 
\def\confName{ICCV}
\def\confYear{2025}

\title{Boosting Vision Semantic Density with Anatomy Normality Modeling for Medical Vision-language Pre-training}

\author{Weiwei Cao$^{1,2,4}$ \quad Jianpeng Zhang$^{1,2,4}$\thanks{Correspondence to Jianpeng Zhang.} \quad Zhongyi Shui$^{2}$ \quad Sinuo Wang$^{2}$ \quad Zeli Chen$^{2}$ \\ Xi Li$^{1}$ \quad Le Lu$^{2}$ \quad Xianghua Ye$^{3}$ \quad Tingbo Liang$^{3}$ \quad Qi Zhang$^{3}$ \quad Ling Zhang$^{2}$\\
\normalsize{\textsuperscript{1}College of Computer Science and Technology, Zhejiang University} \quad
\normalsize{\textsuperscript{2}DAMO Academy, Alibaba Group} \\
\normalsize{\textsuperscript{3}The First Affiliated Hospital of College of Medicine, Zhejiang University} \quad
\normalsize{\textsuperscript{4}Hupan Lab, 310023, Hangzhou, China} \\
{\tt\small {jianpeng.zhang0@gmail.com}}
}

\begin{document}
\maketitle
\begin{abstract}

\vspace{-0.3cm}
Vision-language pre-training (VLP) has great potential for developing multifunctional and general medical diagnostic capabilities.  
However, aligning medical images with a low signal-to-noise ratio (SNR) to reports with a high SNR presents a semantic density gap, leading to visual alignment bias.
In this paper, we propose boosting vision semantic density to improve alignment effectiveness. 
On one hand, we enhance visual semantics through disease-level vision contrastive learning, which strengthens the model's ability to differentiate between normal and abnormal samples for each anatomical structure. 
On the other hand, we introduce an anatomical normality modeling method to model the distribution of normal samples for each anatomy, leveraging VQ-VAE for reconstructing normal vision embeddings in the latent space. This process amplifies abnormal signals by leveraging distribution shifts in abnormal samples, enhancing the model's perception and discrimination of abnormal attributes. 
The enhanced visual representation effectively captures the diagnostic-relevant semantics, facilitating more efficient and accurate alignment with the diagnostic report. 
We conduct extensive experiments on two chest CT datasets, CT-RATE and Rad-ChestCT, and an abdominal CT dataset, MedVL-CT69K, 
and comprehensively evaluate the diagnosis performance across multiple tasks in the chest and abdominal CT scenarios, achieving state-of-the-art zero-shot performance. Notably, our method achieved an average AUC of 84.9\% across 54 diseases in 15 organs, significantly surpassing existing methods. Additionally, we demonstrate the superior transfer learning capabilities of our pre-trained model. 
Code is available at 
\def\UrlFont{\rm\small\ttfamily}
\url{https://github.com/alibaba-damo-academy/ViSD-Boost}

\end{abstract}    
\vspace{-0.5cm}
\section{Introduction}
\vspace{-0.2cm}
\label{sec:intro}

\label{sec:intro}

The advancement of computer-aided diagnosis has conventionally depended on supervised learning methodologies that necessitate pixel-level, region-level, or image-level annotations~\cite{isensee2021nnu,seg_review,cls_review}. This process is both time-intensive and labor-intensive, thereby complicating the creation of adaptable and versatile generalist models.
Vision-language pre-training (VLP), driven by natural language, eliminates the need for excessive manual annotation and has achieved significant success in natural image scenarios~\cite{clip,flip,albef}. This approach has the potential to disrupt the traditional supervised learning pipeline, enabling the development of more versatile diagnostic capabilities at a lower cost.

\begin{figure}
    \centering
    \includegraphics[width=1.0\linewidth]{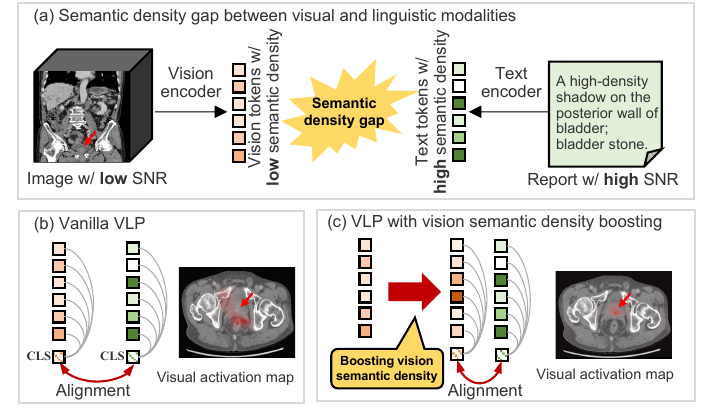}
    \caption{(a) Illustration of semantic density gap between vision and linguistic modalities in the medical scenario. We present an abdomen CT scan, accompanied by the diagnostic report indicating the presence of a bladder stone. SNR: signal-to-noise ratio. (b) In the vanilla VLP, the visual activation map fails to highlight regions of interest for bladder stone diagnosis, resulting in \textit{\textbf{visual alignment bias}}. 
    (c) Our method is proposed to enhance attention to disease-related regions by boosting vision semantic density. 
    }
    \label{fig:idea}
\end{figure}

However, recent attempts in medical scenarios have yielded only modest success, with diagnostic performance falling short of clinical requirements~\cite{yan2023multimodal}. The core challenge lies in extracting diagnostic-related semantics from vision embedding space. 
Medical images encompass a broad range of anatomical content, yet the content of interest relevant to diagnostic decisions is often sparse, potentially occupying only a small portion of the whole image.  
It is difficult to identify diagnostic-related visual cues from a vast amount of image space due to the relatively low signal-to-noise ratio (SNR). 
Here, we introduce the concept of \textbf{\textit{semantic density}}, which specifically refers to the concentration of diagnostic-related signals conveyed within the representation of medical images. 
In conditions of low SNR, diagnostic information may be diluted by a large amount of noise, resulting in a low visual semantic density. 
In contrast, diagnostic reports provide a highly condensed summary of image observations, leading to rich diagnostic-related semantics.
When aligning these two modalities, the gap in semantic density may lead to visual alignment bias. As illustrated in Fig.~\ref{fig:idea}, we present an example of diagnosing a bladder stone in a CT scan. The low visual semantic density hinders VLP from accurately focusing on the small bladder stone, which occupies less than one-thousandth of the whole volume. 

In this paper, we propose to tackle the visual alignment bias by \textbf{Boost}ing \textbf{Vi}sion \textbf{S}emantic \textbf{D}ensity (ViSD-Boost) for medical vision-language pre-training. Our method consists of two key steps. 
(1) \textbf{\textit{Enhancing vision semantics}}: 
We begin by defining ``normal" as the healthy state of an organ, while ``abnormal" refers to the symptom changes resulting from certain diseases. 
We enhance the discrimination of normal and abnormal organs by visual contrastive learning. 
Before that, we prompt the Large Language Model to automatically extract anatomical abnormality labels. For each organ, all samples are categorized into normal and abnormal groups based on the diagnostic description in the report. Our objective is to establish a visual representation distribution such that normal samples of the same organ are semantically similar in the embedding space, while abnormal samples not only deviate from the normal samples but also maintain distinct differences from each other. This is mainly due to the fact that there are no identical patients who differ more or less in lesion size, location, attributes, and pathological types, and recognizing these differences is crucial for semantic understanding.
(2) \textbf{\textit{Increasing vision semantic density}}: Ideally, the visual representation should adequately represent the content relevant to the diagnosis, which necessitates the model to be able to extract disease-related cues from large amounts of visual volume. To enhance the model's ability to capture visual anomalies, we introduce an anatomical normality modeling method to characterize the normal distribution of each anatomy. Specifically, 
we design a lightweight VQ-VAE~\cite{vqvae} that learns the normal distribution from a large number of healthy samples in the latent space. 
Given that abnormal samples exhibit distribution shifts, we can enhance the abnormal components derived from the reconstruction errors, as these components are often closely linked to the diagnosis.

We conduct experiments on chest CT VLP benchmark datasets, CT-RATE~\cite{hamamci2024foundation} and RAD-ChestCT~\cite{draelos2021machine}, and abdomen CT VLP benchmark dataset MedVL-CT69K~\cite{shui2025large}. 
Experimental results indicate that our method outperforms recent state-of-the-art VLP methods, especially in abdominal scenarios, achieving an AUC of 84.9\% in zero-shot diagnostic tasks covering 54 diseases across 15 organs. 
Moreover, our pre-trained model is also superior in several downstream tasks, including radiology report generation, and supervised multi-disease classification. 
Our contributions are summarized as follows:
\begin{itemize}
    \item We introduce the concept of semantic density in medical vision-language scenarios and propose a vision semantic density boosting method to address the visual alignment bias. 
    \item We introduce disease-level contrastive learning to enhance vision semantics for distinguishing normal and abnormal anatomies. 
    \item We propose anatomical normality modeling to establish normal distributions of healthy anatomies and capture abnormal visual cues under distribution shifts, thereby increasing visual semantic density. 

\end{itemize}
\section{Related work}
\label{sec:related_work}

\begin{figure*}[t]
    \centering
    \includegraphics[width=1.0\linewidth]{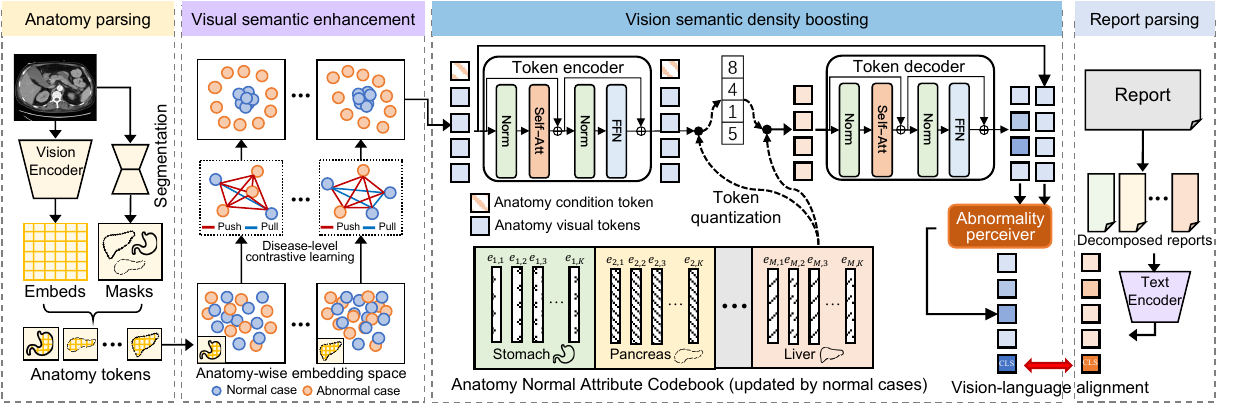}
    \vspace{-0.4cm}
    \caption{The framework of the proposed ViSD-Boost. \textbf{Anatomy parsing}: Extracting anatomical vision tokens based on the segmentation mask; \textbf{Visual semantic enhancement}: Using disease-level contrastive learning to enhance the semantic discrimination between normal and abnormal samples; \textbf{Vision semantic density boosting}: Modeling the distribution of normal samples for each anatomy using VQ-VAE to amplify abnormal vision cues; \textbf{Report parsing}: Decomposing the original report using LLM to generate anatomical-specific reports. }
    \vspace{-0.3cm}
    \label{fig:norm_learning}
\end{figure*}

\subsection{Medical vision-language learning}
Recently, the advent of visual-language models has provided new avenues for supervised learning~\cite{clip,flip,glip,uniter}. 
The fundamental concept behind these methods is to employ vision and language contrastive learning to align different modalities within the same representation space. 
In the medical domain, several studies have applied contrastive learning to align 2D X-ray images with their corresponding reports, yielding promising outcomes in diverse scenarios~\cite{chexzero,cheng2023prior,zhou2023advancing,chen2022align_knowledge}. To strengthen the alignment, some works integrate local alignment into global contrastive learning. Notable methods such as GLoRIA~\cite{huang2021gloria}, LoVT~\cite{lovt}, and MGCA~\cite{mgca} have introduced techniques that facilitate the alignment of localized image regions with report sentences. Additionally, some studies have attempted to enhance image-report alignment by incorporating medical knowledge~\cite{DCL,KDA2023,medklip,zhang2020radiology,liu2024bootstrapping}. For instance, Li \etal~\cite{DCL} proposes a dynamic knowledge graph to improve visual and linguistic congruence. Despite the advancements, most of the research has predominantly focused on 2D X-ray images. Recently, some works have also begun to explore 3D CT visual-language learning~\cite{biud,ct-glip,hamamci2024foundation,merlin,shui2025large}. These efforts demonstrate, to some extent, the potential of vision-language learning in 3D image analysis. 
However, these methods still have not overcome the bottleneck of vision semantic density, making it difficult to extract diagnostic-related visual cues in the complex 3D abdomen scenario. This may also explain why most attempts in the field are still limited to relatively simple 2D chest scenarios.

\subsection{Visual representation enhancement}

Visual representation learning has long been a research hotspot in the computer vision community and is also critical for medical image analysis~\cite{lovt,mgca}. 
Recently, some works have utilized visual representation learning to enhance vision semantics in VLP~\cite{zhou2023advancing,biud}.
Most methods can be categorized into two paradigms, \ie, supervised learning and self-supervised learning. 
Supervised learning typically involves training a vision encoder on labeled data, which is then transferred to VLP~\cite{li2024organ,asg,zhou2024large}. For example, Li \etal~\cite{asg} introduced an additional disease classification task to the visual model, improving its ability to identify fourteen thoracic abnormalities. However, this approach is prone to overfitting on specific labeled categories, leading to a lack of generalization in representations.
In contrast, self-supervised learning does not require any labeled data~\cite{zhou2023advancing,flip}. Instead, it learns visual representations through pretext tasks, e.g., contrastive learning~\cite{chen2020simple}, or masked image modeling~\cite{he2022masked}. The advantage of these methods is that the representations are sufficiently general, but the model primarily focuses on instance-level representation learning, lacking disease-level semantics.
In contrast to these approaches, we propose a disease-level visual representation learning strategy to enhance the representation capability of vision semantics, ensuring it is both sufficiently general and enriched with disease-specific semantics. 
\section{Approach}
\label{sec:approach}

\subsection{Anatomy-wise image-report alignment}

We denote a dataset with paired image and report as $\{X_i^I, X_i^R; i=1,...,N\}$. 
Following~\cite {shui2025large}, we decompose the image and report of each paired data based on anatomical units. 
First, we parse the segmentation structure of each organ by using a whole-body segmentation model~\cite{totalsegmentator}, $X_i^I \to \{X_{i,j}^I;j=1,...,M\}$, where $M$ is the number of anatomical structures. 
Second, we utilize Qwen~\cite{bai2023qwen} to decompose the diagnostic report into a structured report at the anatomical level, $X_i^R \to \{X_{i,j}^R;j=1,...,M\}$.

Considering the superior capability of convolution in the local feature extraction, we utilize the residual convolutional network~\cite{he2016deep} as the vision encoder to extract the anatomical vision feature map $f_{i,j}^I$. Subsequently, the vision feature is flattened along the spatial dimension to obtain a sequence of anatomical visual tokens. 
For the text encoder, we employ a pre-trained Bert model to extract anatomical-level report tokens $f_{i,j}^R$. Additionally, we append learnable query tokens specific to the visual and textual tokens for each anatomy to aggregate all tokens via the cross attention, denoted as $Q_{i,j}^{I} = {\color{gray}\texttt{CrossAttn}}(Q_{i,j}^{I}, f_{i,j}^{I}, f_{i,j}^{I})$ and $Q_{i,j}^{R} = {\color{gray}\texttt{CrossAttn}}(Q_{i,j}^{R}, f_{i,j}^{R}, f_{i,j}^{R})$.  
Overall, the learning objective of vision-language pre-training can be formulated as 
\begin{equation}
\small
\label{eq.VLP}
    \underset{\theta^I,\theta^R}{\arg\min} -\frac{1}{B * M}\sum_{i=1}^{B} \sum_{j=1}^{M}\log(\frac{e^{\langle {Q}_{i,j}^{I}, {Q}_{i,j}^{R} \rangle /\tau}}{\sum_{k=1}^{B} e^{\langle {Q}_{i,j}^{I}, {Q}_{k,j}^{R} \rangle /\tau}})
\end{equation}
where $\theta^I,\theta^R$ are the parameters of the vision and text encoder, $B$ is the number of samples in a mini-batch, and $\tau$ is the temperature.

\subsection{Visual semantic enhancement}

We introduce a visual semantic enhancement method that employs disease-level contrastive learning to improve the discrimination ability between normal and abnormal anatomies. 
Typically, conventional visual contrastive learning focuses primarily on instance-level representation, wherein samples are pushed away from each another~\cite{oord2018representation,he2020momentum}. However, such an approach is inadequate for obtaining disease-level semantics. 
Inspired by anomaly detection~\cite{pang2021deep}, we anticipate that normal and abnormal samples should exhibit a distinctive representation distribution within the embedding space. In particular, normal samples belong to the same category, and consequently, their representations should cluster closely in the embedding space, extending beyond simply consistent views of the same instance. 
Conversely, abnormal samples should not be considered as belonging to the same category. These abnormal organs are unlikely to exhibit identical abnormal characteristics, as they may differ in lesion location, size, shape, pathological type, and other factors. 
Distinguishing these abnormal instances enhances the model's ability to comprehend their unique characteristics, thereby improving semantic understanding.

Given a batch of $B$ paired image and report samples, we first utilize the diagnostic reports to determine the status of each anatomical structure, i.e., healthy or sick. Specifically, we leverage prompt learning to enable Qwen~\cite{bai2023qwen} to analyze the description of each organ mentioned in the report. Organs assessed as completely healthy are classified as normal, while any identified abnormalities are classified as abnormal. The resulting organ-level abnormality labels are defined as $y \in \{0:\texttt{normal}, 1:\texttt{abnormal}\}^{B\times M}$. 
We then design the following disease-level contrastive learning loss to optimize the representation distribution for each anatomical structure, expressed as
\begin{equation}
\begin{aligned}
\tiny
    -\frac{1}{B * M}\sum_{i=1}^{B} \sum_{j=1}^{M} \{\mathbb{I}_{y_{i,j}=1} \log(\frac{e^{\langle {Q}_{i,j}^{I}, {Q}_{i,j}^{I'} \rangle /\tau}}{\sum_{k=1}^{B} e^{\langle {Q}_{i,j}^{I}, {Q}_{k,j}^{I'} \rangle /\tau}}) \\
    + \sum_{p=1}^{B} \mathbb{I}_{y_{i,j}=0} \mathbb{I}_{y_{p,j}=0} \log(\frac{e^{\langle {Q}_{i,j}^{I}, {Q}_{p,j}^{I'} \rangle /\tau}}{\sum_{k=1}^{B} e^{\langle {Q}_{i,j}^{I}, {Q}_{k,j}^{I'} \rangle /\tau}})  \}
\end{aligned}
\end{equation}
where $\mathbb{I}$ is the indicator function.
To avoid a collapsed solution for the abnormal cases, such as matching the same vector $Q_{i,j}^I$ for the same instance $X_{i,j}^I$, we first employ data augmentation to construct different views $X_{i,j}^{I'}$, and then maintain a slow-moving average vision encoder (momentum encoder) to generate $Q_{i,j}^{I'}$ as the positive pair for $Q_{i,j}^I$. 
It is noteworthy that this representation learning process is performed before the vision-language pre-training, with specific training details introduced in Sec.~\ref{sec.impl_details}.

\subsection{Vision semantic density boosting}
\subsubsection{Anatomical normality modeling}

After the semantic enhancement, the vision encoder has gained the ability to distinguish between normal and abnormal samples. However, its capacity to capture critical diagnostic-related cues from the whole anatomical region remains insufficient. Here, we introduce an approach called anatomical normality modeling, utilizing Vector Quantised Variational AutoEncoder (VQ-VAE)~\cite{van2017neural} to learn the normal distribution of healthy anatomical structures.
Our approach differs from conventional VQ-VAE in the following two aspects: 1. \textbf{\textit{Multi-distribution learning}}. In our scenario, CT images encompass dozens of anatomical structures, necessitating simultaneous normality modeling for multiple anatomies. Therefore, we specifically introduce an anatomical condition token for each anatomy, prompting the VQ-VAE to perform the reconstruction task for a specific anatomy. 2. \textbf{\textit{Modeling in latent space}}. We train the VQ-VAE in the latent space rather than the image space, which not only enhances computational efficiency but also facilitates the encoding of normality attributes in the high-level semantic space. 
Specifically, we design a Transformer-based~\cite{vaswani2017attention} token encoder $\varphi_E$ and token decoder $\varphi_D$ as the backbone of the VQ-VAE. The token encoder encodes the anatomical tokens into a discrete codebook space, and subsequently, the nearest-neighbor vectors from the codebook are utilized to reconstruct the tokens via the token decoder. The design of the encoder network is crucial for the construction of the codebook. The codebook should represent the multifaceted and rich attributes of normal anatomies, such as organ shape, texture, and intensity, which require global aggregation of whole organ tokens and are not suitable for locality encoding typically performed by convolutional neural networks~\cite{guo2022cmt}. Consequently, we utilize Transformers to build the token encoder and decoder, as they are better equipped to model the long-range dependencies among tokens.
The discrete codebook, composed of $M*K$ vectors, is defined as $e \in \mathbb{R}^{M \times K \times C}$. Here \textit{M} represents the number of anatomical structures, \textit{K} denotes the number of prototype vectors set for each anatomy, and \textit{C} is the dimensionality of the vectors. 
The codebook is slowly updated with normal embeddings via the exponential moving average strategy~\cite{van2017neural} during the training process.
The learning process for embedding reconstruction can be mathematically formulated as follows 
\begin{equation}
\begin{aligned}
\tiny
    -\frac{1}{B * M}\sum_{i=1}^{B} \sum_{j=1}^{M} & {\mathbb{I}_{y_{i,j}=0}} \cdot \{ \left \| f_{i,j}^I - {\varphi_D}(e_{j,k}) \right\|_2^2  \\
    &+ \beta  \left \| \texttt{sg}[e_{j,k}] - \varphi_E(f_{i,j}^I;A_j) \right\|_2^2 \}
\end{aligned}
\end{equation}
where $k = \mathop{\arg\min}\limits_{m} \left\| {\varphi_E}(f_{i,j}^I;A_j) - e_{m} \right\|_2^2$, $A_j$ is the anatomy condition token, $\beta$ is the weight balancing factor with a default setting of 0.25, and $\texttt{sg}$ refers to the stop gradient operation.
Once trained, the model will exhibit diminished reconstruction quality when handling abnormal data, as this type of data generally deviates from the normal distribution. As a result, low-quality reconstructions can be viewed as indicators of abnormality, thereby achieving our objective of detecting abnormal signals.

\subsubsection{Abnormality semantic perception}

Let $q_{i,j}^I$ represents the reconstructed embedding of the original embedding $f_{i,j}^I$ by the VQ-VAE.
It is important to note that this embedding does not present any abnormal semantics. It is necessary to design a discrepancy-aware perception module that, using the reconstructed normal embedding as a reference, can detect differences in the original embedding, indicating potential abnormal components. The module is expected to extract and amplify those signals to enhance the semantic density of the vision embedding. 
To this end, we introduce a simple yet efficient perception module that concatenates $f_{i,j}^I$ and $q_{i,j}^I$ as input to a multiple-layer perceptron (MLP) network. We replace the original embedding $f_{i,j}^I$ with the output of MLP, denoted as $\widehat{f}_{i,j}^I$, and perform the vision-language pre-training according to Eq~\ref{eq.VLP}.

\begin{table}[t!]
\centering
\resizebox{\linewidth}{!}{
\begin{tabular}{c|c|cccc}
\toprule
Dataset & Method              & Precision & ACC  & F1 & AUC  \\ \midrule
\multirow{10}{*}{\begin{tabular}[c]{@{}c@{}}Internal \\ validation   \\ (CT-RATE)\end{tabular}}     
& Random~\cite{hamamci2024foundation}              & 18.0      & 50.2 & 57.0     & 50.5 \\
& Supervised~\cite{hamamci2024foundation}          & 24.0      & 58.1 & 63.2     & 60.3 \\
& CT-CLIP~\cite{hamamci2024foundation} & 32.6    & 66.9    & 70.8      & 73.3  \\ 
& CT-VocabFine\textsuperscript{\textdagger}        & 35.6      & 70.4   & 73.8     & 76.0  \\
& CT-LiPro\textsuperscript{\textdagger}            & 34.3  & 69.1   & 72.6     & 76.1  \\ 
& BIUD~\cite{biud} & 33.8      & 68.1 & 71.6 & 71.3 \\
& Merlin~\cite{merlin} & 33.7      & 67.2 & 70.9 & 72.8 \\
& fVLM~\cite{shui2025large} & 37.9      & 71.8 & 75.1 & 77.8 \\ 
& ViSD-Boost   & \textbf{38.7}      & \textbf{73.1} & \textbf{75.9}     & \textbf{79.0} \\ \midrule
\multirow{10}{*}{\begin{tabular}[c]{@{}c@{}}External \\ validation   \\ (Rad-ChestCT)\end{tabular}} 
& Random~\cite{hamamci2024foundation}              & 26.5      & 50.0 & 55.5     & 49.6 \\
& Supervised~\cite{hamamci2024foundation}          & 28.7      & 53.9 & 58.7     & 54.1 \\
& CT-CLIP~\cite{hamamci2024foundation} & 34.1   & 59.9  & 64.7   & 63.2 \\ 
& CT-VocabFine\textsuperscript{\textdagger}        & 35.6    & 62.1  & 66.8    & 65.7 \\
& CT-LiPro\textsuperscript{\textdagger}           & 35.1  & 60.6  & 65.0  & 64.7 \\
& BIUD~\cite{biud} & 33.7 & 60.6 & 65.2 & 62.9 \\
& Merlin~\cite{merlin} & 34.8 & 61.9 & 66.3 & 64.4 \\
& fVLM~\cite{shui2025large} & \textbf{37.4} & 64.7 & 68.8 & 68.0 \\ 
& ViSD-Boost    & 34.2    & \textbf{65.2}    & \textbf{69.3}     & \textbf{69.4} \\ \bottomrule
\end{tabular}}
\caption{Zero-shot performance comparison on the CT-RATE and Rad-ChestCT datasets. \textsuperscript{\textdagger} denotes the improved version of CT-CLIP that was further fine-tuned by supervised learning. Note that the entities of ``lymphadenopathy" and ``medical material" are excluded from the comparison, and the results of CT-CLIP and its variants are drawn from the latest manuscript available on arXiv.} 
\vspace{-0.4cm}
\label{tab:zero_shot_ct_rate}
\end{table}

\begin{table*}[t!]
\small
\centering
\scalebox{0.8}{
\setlength{\tabcolsep}{1.2mm}{
\begin{tabular}{c|ccc|ccc|ccc|ccc|ccc|ccc|ccc|ccc}
\toprule
\multirow{2}{*}{Method} & \multicolumn{3}{c|}{\textbf{Adrenal gland}} & \multicolumn{3}{c|}{\textbf{Bladder}}  & \multicolumn{3}{c|}{\textbf{Colon}}       & \multicolumn{3}{c|}{\textbf{Esophagus}}       & \multicolumn{3}{c|}{\textbf{Gallbladder}} & \multicolumn{3}{c|}{\textbf{Heart}}   & \multicolumn{3}{c|}{\textbf{Kidney}} & \multicolumn{3}{c}{\textbf{Liver}}   \\ 
                                  & SE            & SP           & AUC          & SE           & SP         & AUC        & SE           & SP           & AUC         & SE            & SP            & AUC           & SE    & SP                         & AUC  & SE          & SP          & AUC       & SE         & SP         & AUC        & SE          & SP         & AUC       \\ \midrule
Supervised  & 57.8          & 65.4         & 64.1         & 30.4         & 89.3       & 73.5       & 67.1         & 74.8         & 76.0        & 60.5          & 96.2          & 93.9          & 56.8  & 58.7      & 63.1   & 56.8        & 69.8        & 64.6      & 55.4       & 63.7       & 62.3       & 66.8        & 77.0       & 78.9      \\ \midrule
CLIP~\cite{clip}                              & 66.6          & 55.4         & 63.2         & 57.7         & 67.6       & 65.1       & 64.4         & 63.2         & 65.8        & 65.1          & 78.3          & 67.3          & 55.5  & 59.9                       & 59.5 & 36.2        & 77.1        & 44.1      & 55.5       & 61.6       & 59.9       & 70.8        & 65.4       & 72.4      \\
LOVT~\cite{lovt}                              & 62.2          & 54.7         & 60.6         & 71.4         & 62.3       & 70.9       & 69.6         & 58.5         & 67.5        & 84.2          & 85.1          & 89.3          & 65.2  & 51.8                       & 61.2 & 84.6        & 64.9        & 78.3      & 62.2       & 54.7       & 60.2       & 68.3        & 60.2       & 69.3      \\
MGCA~\cite{mgca}                              & 56.8          & 56.5         & 57.4         & 68.5         & 63.9       & 69.0       & 72.8         & 60.1         & 70.2        & 69.7          & 87.9          & 83.8          & 55.1  & 61.0                       & 62.1 & 77.1        & 67.6        & 74.9      & 58.1       & 56.8       & 59.0       & 66.9        & 66.5       & 71.0      \\
Imitate~\cite{imitate}                           & 64.3          & 55.9         & 60.2         & 72.6         & 67.9       & 74.1       & 69.2         & 60.7         & 68.0        & 98.1          & 89.4          & 95.6          & 60.0  & 59.7                       & 62.5 & 71.5        & 75.1        & 70.7      & 60.7       & 55.8       & 59.9       & 66.5        & 65.4       & 69.8      \\
ASG~\cite{asg}                               & 58.0          & 57.0         & 59.0         & 74.4         & 68.0       & 72.6       & 68.6         & 62.6         & 67.0        & 98.7          & 86.5          & 93.3          & 60.0  & 55.5                       & 58.4 & 68.1        & 74.2        & 69.0      & 58.0       & 57.0       & 59.0       & 66.4        & 66.1       & 70.9      \\
BIUD~\cite{biud}                              & 64.9          & 56.0         & 63.4         & 79.8         & 73.5       & 81.0       & 70.4         & 64.8         & 70.0        & 54.1          & 91.9          & 62.6          & 60.6  & 61.1                       & 64.2 & 68.2        & 56.1        & 62.1      & 60.8       & 61.8       & 63.7       & 72.6        & 74.0       & 79.2      \\
Merlin~\cite{merlin}                            & 58.9          & 57.9         & 60.3         & 70.8         & 73.0       & 76.9       & 71.1         & 62.0         & 69.1        & 41.5          & 88.5          & 49.2          & 64.6  & 53.5                       & 61.2 & 69.6        & 75.1        & 72.8      & 58.6       & 64.5       & 64.2       & 73.6        & 75.9       & 80.1      \\ 
fVLM~\cite{shui2025large} &63.0  &63.9 &65.7 &76.2 &77.3 &84.0 &76.1 &75.1 &80.8 &94.4 &96.1 &98.2 &64.9 &58.8 &64.8 &87.2 &75.8 &85.8 &67.9 &72.5 &74.5 &77.2 &76.0 &82.5  \\ 
ViSD-Boost & 63.5          & 64.9         & 68.5         & 75.0         & 74.4       & 81.2       & 77.6         & 76.3         & 81.9        & 99.4          & 92.7          & 98.3          & 65.6  & 69.7                       & 72.6 & 84.6        & 82.7        & 90.5      & 72.4       & 74.5       & 78.5       & 78.4        & 80.3       & 85.9      \\ \midrule
\multirow{2}{*}{Method}                 & \multicolumn{3}{c|}{\textbf{Lung}}          & \multicolumn{3}{c|}{\textbf{Pancreas}} & \multicolumn{3}{c|}{\textbf{Portal vein}} & \multicolumn{3}{c|}{\textbf{Small Intestine}} & \multicolumn{3}{c|}{\textbf{Spleen}}      & \multicolumn{3}{c|}{\textbf{Stomach}} & \multicolumn{3}{c|}{\textbf{Sacrum}} & \multicolumn{3}{c}{\textbf{Average}} \\
                                  & SE            & SP           & AUC          & SE           & SP         & AUC        & SE           & SP           & AUC         & SE            & SP            & AUC           & SE    & SP                         & AUC  & SE          & SP          & AUC       & SE         & SP         & AUC        & SE          & SP         & AUC       \\ \midrule
Supervised  & 45.8          & 89.0         & 51.5         & 73.1         & 70.5       & 78.3       & 81.7         & 87.8         & 91.0        & 74.2          & 76.1          & 81.3          & 62.2  & 78.2          & 76.1   & 63.3        & 72.6        & 73.6      & 29.4       & 92.8       & 77.1       & 62.0        & 76.2       & 73.3      \\ \midrule
CLIP~\cite{clip}                              & 80.4          & 96.1         & 88.3         & 65.4         & 62.4       & 65.0       & 72.4         & 72.4         & 78.6        & 64.4          & 63.2          & 74.5          & 72.8  & 65.9                       & 71.1 & 62.5        & 68.0        & 68.6      & 47.1       & 56.0       & 47.0       & 65.5        & 68.0       & 68.4      \\
LOVT~\cite{lovt}                              & 78.7          & 65.0         & 80.9         & 68.3         & 62.5       & 67.8       & 82.6         & 60.2         & 75.5        & 72.4          & 61.5          & 70.5          & 70.1  & 49.0                       & 66.1 & 62.9        & 67.9        & 69.1      & 70.6       & 38.8       & 48.9       & 70.8        & 60.1       & 69.4      \\
MGCA~\cite{mgca}                              & 81.4          & 71.5         & 82.9         & 67.9         & 64.6       & 70.3       & 77.1         & 65.3         & 76.5        & 67.7          & 67.8          & 72.1          & 67.2  & 64.0                       & 66.6 & 68.8        & 62.7        & 68.5      & 52.9       & 40.9       & 45.0       & 68.3        & 64.5       & 70.1      \\
Imitate~\cite{imitate}                           & 81.1          & 89.7         & 86.7         & 65.0         & 61.3       & 64.3       & 76.1         & 69.3         & 80.5        & 76.1          & 68.0          & 77.6          & 64.0  & 68.9                       & 71.3 & 64.0        & 63.7        & 66.3      & 35.3       & 43.4       & 29.0       & 69.2        & 66.6       & 70.6      \\
ASG~\cite{asg}                               & 73.6          & 98.1         & 89.6         & 66.8         & 60.6       & 64.8       & 74.3         & 76.6         & 80.5        & 71.1          & 70.0          & 75.1          & 66.3  & 63.3                       & 68.3 & 64.3        & 64.0        & 66.7      & 52.9       & 37.8       & 38.6       & 68.2        & 67.5       & 70.1      \\
BIUD~\cite{biud}                              & 69.3          & 84.5         & 72.1         & 72.4         & 70.3       & 76.9       & 82.5         & 71.4         & 82.2        & 74.8          & 67.5          & 75.1          & 65.6  & 72.3                       & 72.3 & 63.1        & 63.8        & 66.1      & 70.6       & 29.2       & 43.8       & 69.3        & 69.0       & 71.4      \\
Merlin~\cite{merlin}                            & 76.9          & 80.1         & 78.7         & 74.2         & 63.8       & 73.5       & 86.2         & 78.0         & 85.9        & 73.4          & 72.1          & 78.4          & 67.3  & 72.2                       & 72.0 & 63.3        & 67.7        & 69.9      & 47.1       & 65.8       & 48.2       & 69.2        & 69.7       & 71.9      \\ 
fVLM~\cite{shui2025large} &74.3  &78.9  &82.2  &75.8  &80.8  &85.3  &90.8  &93.2  &96.7  &74.0  &78.6  &82.1  &76.5  &78.0  &82.0  &69.9  &67.8  &74.1  &88.2  &83.3  &87.5  &75.8     & 76.5 & 81.3     \\ 
ViSD-Boost                              & 89.4          & 87.8         & 92.3         & 80.7         & 85.1       & 88.9       & 92.7         & 92.7         & 97.3        & 84.1          & 79.1          & 88.3          & 78.2  & 77.1  & 82.9 & 73.1        & 77.4        & 81.1      & 70.6       & 75.7       & 77.5       & \textbf{79.6}        & \textbf{79.4}       & \textbf{84.9}      \\ \bottomrule
\end{tabular}}}
\caption{Zero-shot performance comparison on the MedVL-CT69K test set. The results presented are the average performance across 54 diseases on 15 anatomies. Detailed performances for each disease can be found in the Sup. Mat.~\ref{sup_sec:detail_performance}.}
\vspace{-0.5cm}
\label{tab:zero_shot_medvl}
\end{table*}

\section{Experiments}
\label{sec:experiments}

\subsection{Datasets}

\noindent\textbf{Chest CT scenario.}
We conducted experiments on two public datasets: CT-RATE~\cite{hamamci2024foundation} and RAD-ChestCT~\cite{draelos2021machine}. CT-RATE contains 50,188 chest CT scans from 21,304 patients. Following~\cite{hamamci2024foundation}, we split 20,000 patients as the training set and 1,304 patients as the test set. To evaluate generalizability, RAD-ChestCT, which comprises 3,630 CT volumes, was used as an external test set. 
In this scenario, we trained the model from scratch on the CT-RATE training set and tested it on both the CT-RATE test set and RAD-ChestCT.

\noindent\textbf{Abdomen CT scenario.}
We also conducted experiments on the large-scale abdominal CT dataset, MedVL-CT69K~\cite{shui2025large}, which encompasses 272,124 CT scans from 69,086 patients, along with their corresponding reports. Following~\cite{shui2025large}, we split the dataset into training, validation, and test sets, comprising 64,476, 1,151, and 3,459 patients, respectively. 
We trained the model from scratch on the MedVL-CT69K training set and tested it on the MedVL-CT69K test set. 

\subsection{Implementation details}
\label{sec.impl_details}
\textbf{Data pre-processing}. We utilize the TotalSegmentator~\cite{totalsegmentator} to segment 104 anatomical structures from a CT scan. Considering that the report descriptions may not align with such granular segmentation, we group the 104 anatomical structures into 36 primary anatomies, \ie anatomy number $M=36$, as done in~\cite{shui2025large}. This grouping facilitates a more effective alignment between anatomy image and report. All CT images were resampled to $1 mm\times 1 mm\times 5 mm$, with Hounsfield Unit (HU) values truncated to the range of [-1000, 1000] and subsequently normalized to [0, 1]. Whole CT volumes were randomly cropped with the patch size of $256\times 384 \times 96$ as the model input. During training, we only consider the complete organs within the current patch, ignoring any organ parts that may be incomplete due to the cropping operation. 
\textbf{Training steps} include (1) training the vision encoder by disease-level contrastive learning, (2) performing vision-language alignment; (3) training the VQ-VAE with the frozen vision encoder, and (4) fine-tuning the whole vision-language framework with the frozen VQ-VAE. 
It is important to emphasize that for the CT-RATE dataset, we conduct training of all four phases from scratch without utilizing any data or pretrained weights from MedVL-CT69K. 
\textbf{Architecture details}. The vision encoder is based on a 3D ResNet18. The vector number $K$ for each anatomy in the codebook is 100 and dimensionality $C$ is 1024. For 2D VLP methods used for comparison, such as CLIP~\cite{clip}, LOVT~\cite{lovt}, etc., following ~\cite{hamamci2024foundation,biud, shui2025large}, we replace their 2D vision encoders to 3D versions to accommodate CT volumes.
\textbf{Zero-shot diagnosis}: Following previous work~\cite{hamamci2024foundation}, we perform the zero-shot classification using the pre-trained vision encoder and text encoder. 
\textbf{Radiology report generation}: We integrate the pre-trained vision encoder with an additional text decoder~\cite{li2022blip} to perform the downstream radiology report generation task.
\textbf{Multi-disease classification}: We augment the vision encoder by integrating two additional fully connected layers for downstream multi-disease classification tasks.
\textbf{Evaluation Metrics}: Sensitivity (SE), specificity (SP), and Area Under the Curve (AUC) are used to assess the zero-shot diagnostic performance. 
Precision, Recall, F1-score, GREEN~\cite{ostmeier2024green}, BLEU4, ROUGE-L, METEOR, and CIDEr are used to assess the report generation performance.  
We extract the entities from the generated reports by using a text classifier~\cite{shui2025large}, which is able to accurately identify 54 diseases in reports.

\subsection{Zero-shot diagnosis}
We compare the zero-shot diagnostic performance of different methods on both the internal dataset, CT-RATE, and the external dataset, Rad-ChestCT, in Table~\ref{tab:zero_shot_ct_rate}. 
Our method, VisD-Boost, outperformed these VLM methods, achieving AUC scores of 79.0\% and 69.4\% on the internal and external test sets, respectively.
Notably, methods based on fine-grained alignment (fVLM and ViSD-Boost) exhibit considerable performance advantages over global alignment methods (including CT-CLIP, BIUD, Merlin, etc.). However, we are still able to improve upon the most competitive fVLM method by 1.2\% on the internal test set and 1.4\% on the external test set. Additionally, compared to the fine-tuning versions of CT-CLIP (CT-VocabFine and CT-LiPro), our method does not require any fine-tuning yet maintains superior performance. This further highlights the generalizability and potential of our model in open disease diagnostic scenarios.
In Table~\ref{tab:zero_shot_medvl}, we also evaluate the performance of different models on the larger-scale abdomen benchmark. We compare the diagnostic capabilities of 9 different methods across 54 entities related to 15 organs. We observe that the supervised method performs relatively poorly, lacking significant advantages over VLMs. We believe this may be due to the fact that, despite the large scale of the data, it might not be sufficient for classification tasks, leading to potential overfitting risks for the supervised model. This underscores the clear advantages of vision-language models over supervised models in terms of generalizability and versatile diagnostic abilities.
Additionally, compared to other VLMs, our method achieves an overall AUC of 84.9\%, surpassing the second-best method, fVLM, by 3.6\%. This improvement can largely be attributed to the more accurate vision-language alignment facilitated by the vision semantic boosting strategy.

\begin{table*}[ht!]
\centering
\begin{minipage}{0.59\textwidth}
\centering
\small
\resizebox{0.99\linewidth}{!}{
\setlength{\tabcolsep}{1.5mm}{
\begin{tabular}{c|ccccccccc}
\toprule
Encoder                     & Init       & P & R & F1 & GREEN  & BLEU4 & ROUGE-L & METEOR & CIDEr \\ \midrule
\multirow{7}{*}{Frozen}     & Supervised & 19.1      & 18.6   & 13.2   & 25.9    & 12.8   & 40.6    & 30.8   & 6.6   \\
                            & MAE~\cite{mae}   & 8.9   & 5.9    & 4.3  & 21.6    & 13.1   & 41.6    & 30.5   & 6.1   \\
                            & CLIP~\cite{clip}   & 21.6      & 20.4   & 14.6  & 33.4  & 15.5   & 42.2    & 31.0   & 9.6   \\
                            & BIUD~\cite{biud}     & 17.0      & 21.4   & 15.9  & 33.7   & 18.9   & 44.2    & 29.1   & 13.9  \\
                            & Merlin~\cite{merlin}   & 22.6      & 20.9   & 20.7  & 34.2   & 19.0   & 43.8    & 30.0   & 14.3  \\
                            & fVLM~\cite{shui2025large}  & 24.0   & 31.6  & 26.5  & 37.2   & 19.6   & 45.1    & 31.3   & 14.9  \\
                            & ViSD-Boost       & \textbf{34.3}      & \textbf{39.3}  & \textbf{35.2}  & \textbf{44.4}   & \textbf{24.7}   & \textbf{48.7}    & \textbf{32.7}   & \textbf{27.3}  \\ \midrule
\multirow{7}{*}{Finetuning} & Supervised & 18.0      & 28.3   & 20.4  & 35.5   & 17.9   & 43.4    & 30.6   & 11.7  \\
                            & MAE~\cite{mae}        & 13.4      & 14.1   & 10.9  & 29.4   & 15.1   & 42.5    & 30.3   & 8.8   \\
                            & CLIP~\cite{clip}       & 21.0      & 29.5   & 23.2  & 37.6   & 19.5   & 44.8    & 30.7   & 14.3  \\
                            & BIUD~\cite{biud}     & 26.1      & 31.6   & 24.2  & 38.8   & 19.0   & 44.7   & 30.9   & 13.9  \\
                            & Merlin~\cite{merlin}    & 27.5     & 29.9   & 25.8  & 39.2   & 20.9   & 46.0   & 31.1   & 17.2  \\
                            & fVLM~\cite{shui2025large}  & 38.6   & 36.9   & 32.7  & 40.2   & 21.9   & 46.4   & 31.6  & 17.1  \\
                            & ViSD-Boost       & \textbf{39.8}      & \textbf{44.1}   & \textbf{40.9}  & \textbf{46.7}   & \textbf{28.4}   & \textbf{51.0}    & \textbf{34.1}   & \textbf{50.7}  \\ \bottomrule
\end{tabular}}}
\caption{Radiology report generation performance comparison on the MedVL-CT69K test set. Both the MAE and ``Supervised" are 3D models pre-trained using the MedVL-CT69K training set. The term ``Supervised" refers to a supervised classification model trained specifically on 54 diseases. P: Precision, R: Recall.}
\label{tab:report_generation}
\end{minipage}\hfill
\begin{minipage}{0.39\textwidth}
\centering
\small
\resizebox{0.99\linewidth}{!}{
\setlength{\tabcolsep}{1.8mm}{
\begin{tabular}{c|cc|cc|cc}
\toprule
\multirow{2}{*}{Diseases}           & \multicolumn{2}{c|}{SE} & \multicolumn{2}{c|}{SP} & \multicolumn{2}{c}{AUC} \\
                                    & CLIP       & Ours       & CLIP       & Ours       & CLIP       & Ours       \\ \midrule
Cirrhosis                           & 80.6       & 92.2       & 79.2       & 90.3       & 88.9       & 96.8       \\
Fatty Liver                           & 80.7       & 87.8       & 79.3       & 87.5       & 88.3       & 95.0       \\
Abscess                             & 33.3       & 58.3       & 86.1       & 83.3       & 77.8       & 81.6       \\
Cancer                              & 56.7       & 90.0       & 79.3       & 84.6       & 74.6       & 94.3       \\
GCE                                 & 71.5       & 83.8       & 78.8       & 83.2       & 83.2       & 91.1       \\
Metastase                           & 64.8       & 79.5       & 66.7       & 75.9       & 70.5       & 86.6       \\
IBDD                                & 69.4       & 83.7       & 64.2       & 77.4       & 71.9       & 87.6       \\
Cyst                                & 57.1       & 65.6       & 60.1       & 64.8       & 62.2       & 71.7       \\ \midrule 
Average                             & 64.3       & \textbf{80.9}    & 74.2    & \textbf{80.1}    & 77.2    & \textbf{88.1}       \\ \bottomrule
\end{tabular}}}
\caption{Performance comparison of liver-related multi-diseases classification on the MedVL-CT69K test set. GCE: Glisson’s Capsule Effusion, IBDD: Intrahepatic Bile Duct Dilatation.}
\label{tab:liver_cls}
\end{minipage}
\end{table*}

\subsection{Radiology report generation}
We integrate the pre-trained vision encoder with an additional text decoder~\cite{li2022blip} to generate radiology reports. We conduct experiments with two configurations: one with a frozen vision encoder and the other with a fine-tuning vision encoder. 
For comparison, we include two methods of visual semantic enhancement.
The first approach involves enhancing visual representations through self-supervised learning using masked image modeling~\cite{mae}. The second method employs supervised classification learning with disease labels to boost visual representations. 
Additionally, we also compare approaches based on contrastive learning and 3D CT VLP methods, \ie CLIP~\cite{clip}, BIUD~\cite{biud}, Merlin~\cite{merlin}, and fVLM~\cite{shui2025large}.
In Table~\ref{tab:report_generation}, 
ViSD-Boost demonstrates a clear advantage across multiple evaluation metrics of report generation. In both the frozen and finetuning settings, compared to other baseline models, ViSD-Boost achieves the highest scores on clinical efficacy metrics (Precision, Recall, F1, Green) and natural language metrics (BLEU4, ROUGE-L, METEOR, CIDEr). For example, in the finetuning mode, the significant improvements in metrics such as F1 and CIDEr indicate that ViSD-Boost has better learned the relationship between images and text, generating reports that not only more accurately reflect disease conditions but also offer higher readability and information completeness.

\begin{figure}[t]
    \centering
    \includegraphics[width=0.9\linewidth]{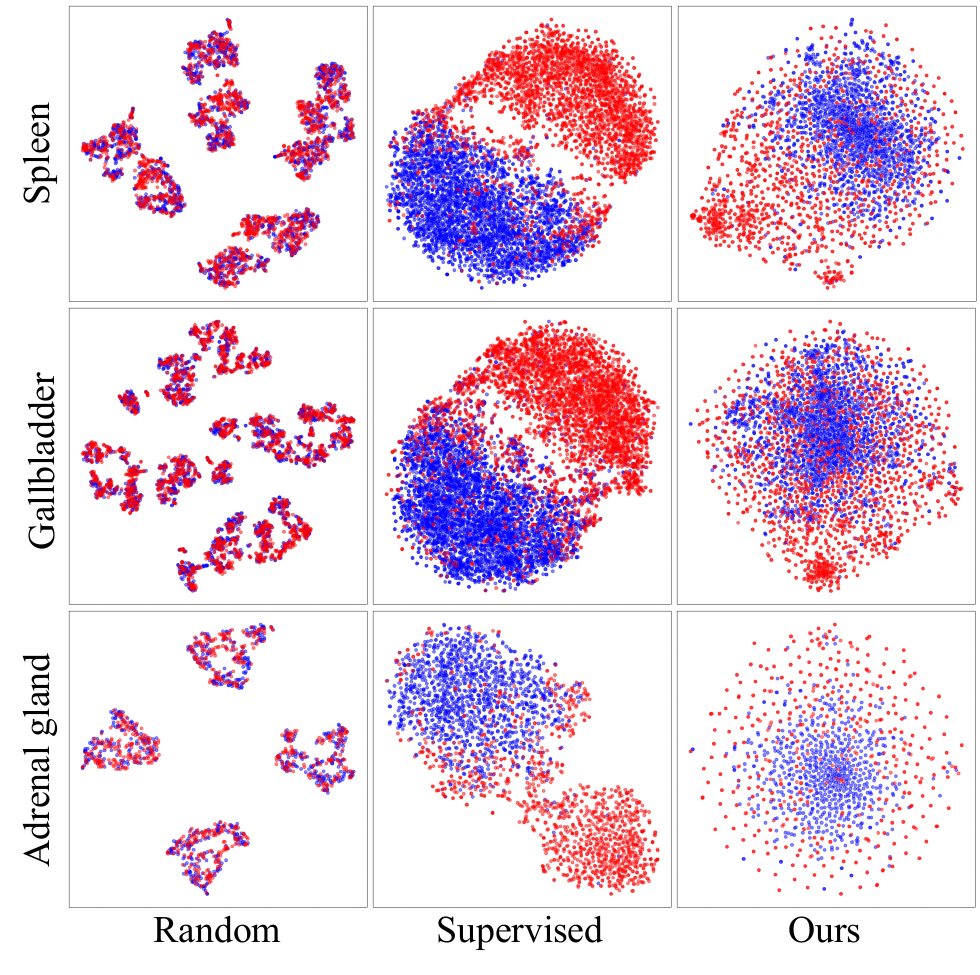}
    \vspace{-0.4cm}
    \caption{T-SNE visualization of normal (blue) and abnormal (red) anatomy embeddings from different methods. Our method is motivated by anomaly detection principles to model the distribution of normal data while promoting variability among normal samples, such as subtle differences in organ size and shape. These variations do not compromise the detection of abnormalities, as the distinction between normal and abnormal samples remains significantly larger. 
}
    \vspace{-0.5cm}
    \label{fig:pretrain_vis}
\end{figure}

\begin{figure*}[t!]
    \centering
    \includegraphics[width=0.9\linewidth]{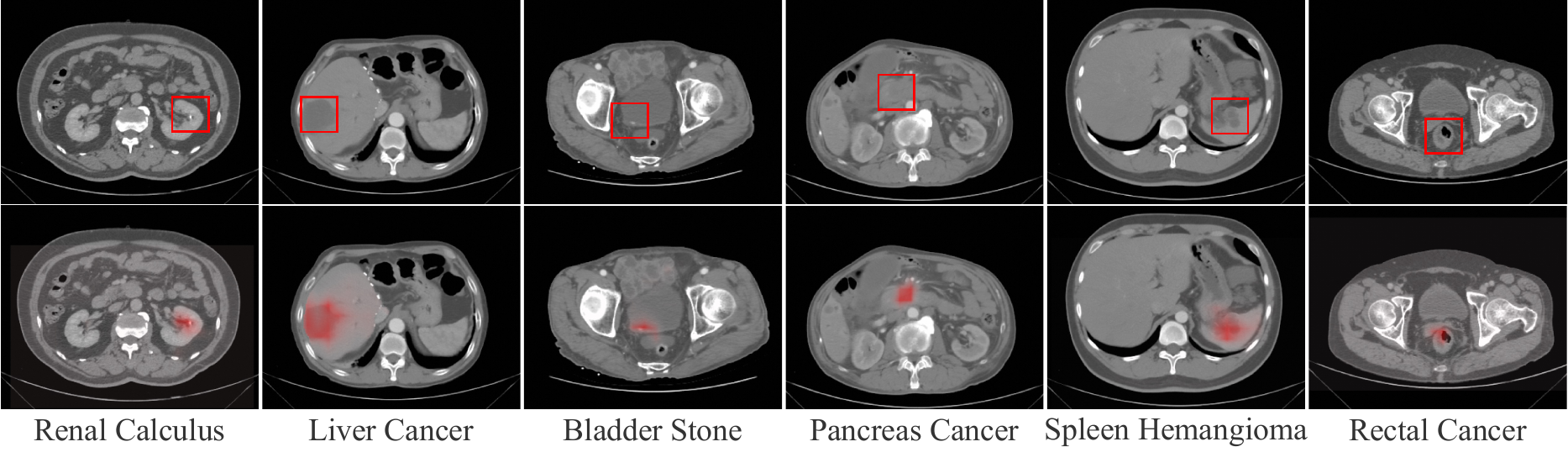}
    \caption{Visual activation maps generated by the proposed method for diagnosing six different diseases.}
    \vspace{-0.5cm}
    \label{fig:attn_map}
\end{figure*}

\subsection{Multi-disease classification}
We conduct a linear probing experiment for multi-disease classification to evaluate the semantic perception capability of the pre-trained vision encoder. To this end, we add a classification head to the vision encoder. Using the text classifier, we have extracted eight liver disease labels from the MedVL-CT69K training set and show the linear probing performance of the MedVL-CT69K test set in Table~\ref{tab:liver_cls}. Compared to CLIP, our method demonstrates the most significant improvement in sensitivity, with an increase of 16.6\%. 
This suggests that the visual representations derived from our model possess greater semantic density, making them effective when transferred to downstream tasks.

\begin{table}[]
\small
\centering
\begin{tabular}{ccccc|cc}
    \toprule
    AAV         & AAC       & VSEI      & VSED      &VSDB       &ACC    & AUC \\ \midrule
                &           &           &           &           &69.3   & 70.9 \\
    \checkmark  &           &           &           &           &73.1   & 76.5 \\
                &\checkmark &           &           &           &74.8   & 78.7 \\
                &\checkmark &\checkmark &           &           &77.3   & 79.7 \\
                &\checkmark &           &\checkmark &           &77.3   & 80.7 \\
                &\checkmark &           &\checkmark &\checkmark &78.0   & 82.5 \\
    \bottomrule
\end{tabular}
\caption{The ablation study of proposed components on MedVL-CT69K validation set. AAV/AAC: Anatomy-wise image-report Alignment with ViT/CNN vision encoder; VSEI/D: Visual Semantic Enhancement with Instance/Disease-level contrastive learning; VSDB: Vision Semantic Density Boosting.}
\vspace{-0.6cm}
\label{tab:ablation}
\end{table}

\subsection{Ablation study}
\subsubsection{Quantitative analysis}
We evaluated the effectiveness of our proposed modules on zero-shot tasks using the MedVL-CT69K validation set. As shown in Table~\ref{tab:ablation}, without fine-grained alignment (first row in the table), performance is poor with an AUC of only 70.9\%. By employing the fine-grained strategy proposed in fVLM, the AUC can be significantly boosted to 76.5\%, which serves as the baseline aligned with fVLM. Our experiments reveal that, first, using a CNN (AAC) as the vision encoder yields superior performance in CT image understanding compared to ViT (AAV), as it is more effective at capturing fine-grained details. Second, disease-level contrastive learning (VSED) delivers stronger diagnostic performance than instance-level contrastive learning (VSEI), demonstrating its effectiveness in enhancing visual semantics. Finally, our proposed vision semantic density boosting (VSDB) further elevates performance on this strong baseline to 82.5\%, representing a 6\% improvement over the baseline of 76.5\%.

\subsubsection{Qualitative analysis}
\noindent \textbf{Visual semantic enhancement}: We conduct an in-depth exploration of the effects of the proposed visual semantic enhancement by T-SNE visualizations, as shown in Figure~\ref{fig:pretrain_vis}. 
To facilitate comparative analysis, we have included two alternative approaches: one using a randomly initialized model and the other employing a supervised classification model that classifies each anatomy as either normal or abnormal.
With random initialization, embeddings of normal and abnormal anatomies are intermixed. 
The supervised classification model creates separation between normal and abnormal anatomies but enforces the abnormal anatomies to group in one single cluster, which diminishes the fine-grained distinctiveness among different types of abnormalities. We do not endorse compact representations like those in supervised classification, as they can oversimplify features, leading to a loss of fine-grained representation and a risk of overfitting. 
In contrast, our proposed method promotes a distribution pattern in which normal samples are clustered together while various abnormal samples remain different from each other. 
This distribution aligns naturally with vision-language pre-training objectives by emphasizing semantic coherence.

\begin{figure}[t!]
    \centering
    \includegraphics[width=1.0\linewidth]{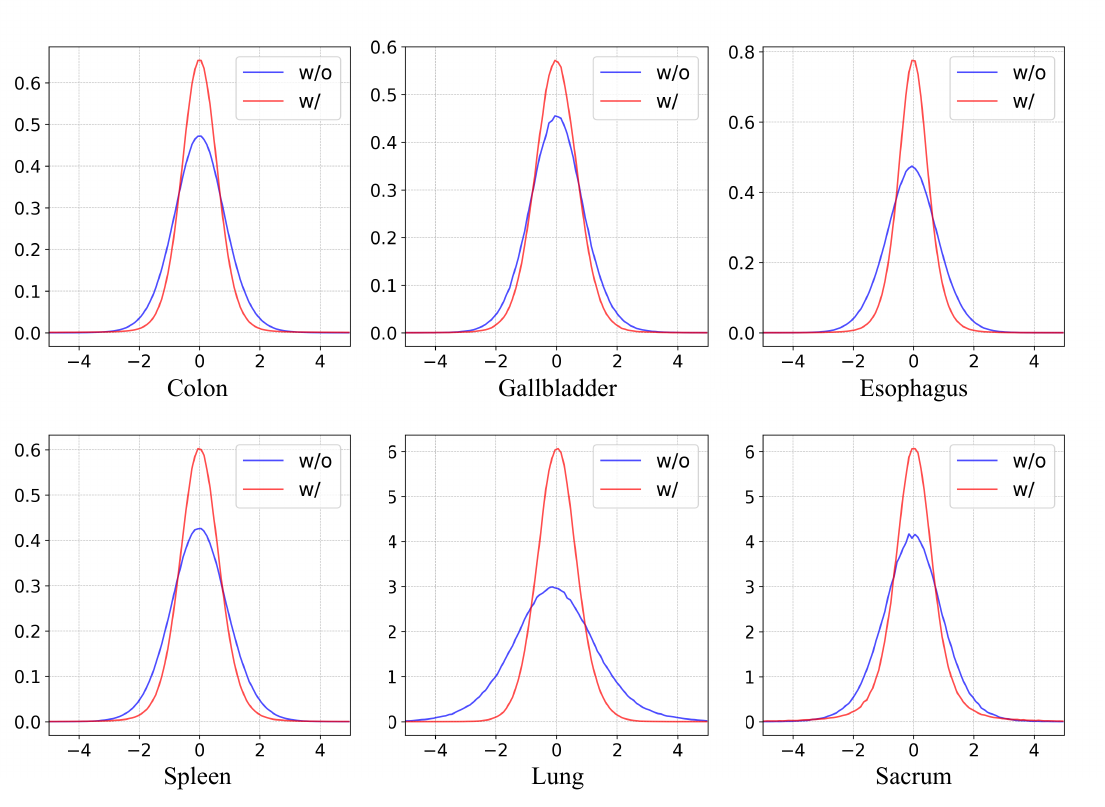}
    \vspace{-0.8cm}
    \caption{Vision semantic density comparison between models w/ and w/o VSDB. The X-axis represents the activation values within the vision tokens, while the Y-axis indicates the frequency.}
    \vspace{-0.5cm}
    \label{fig:density_plot}
\end{figure}

\noindent \textbf{Vision semantic density boosting}: We further demonstrate the effectiveness of the proposed VSDB in improving the model's capability to perceive and capture disease cues through visual activation maps. Figure~\ref{fig:attn_map} visualizes the activation maps highlighting the image regions associated with various diseases. 
To further analyze the effectiveness of VSDB, we observe the distribution of vision tokens obtained by two model variants (w/ and w/o VSDB module). We display the distribution of vision tokens across six anatomical structures in Figure~\ref{fig:density_plot}, with more anatomies shown in Sup. Mat. 
We can see that after employing VSDB, more activations of the vision tokens are concentrated near zero, resulting in a sparser activation. This implies that the overall representation becomes sparser, which encourages the model to focus more on important features, thereby enhancing its semantics.
\section{Conclusion}
\label{sec:conclusion}

In this work, we propose boosting vision semantic density to address visual alignment bias caused by the semantic density gap between medical images and diagnostic reports. 
On one hand, we propose a disease-level visual contrastive learning method to enhance visual semantics. 
On the other hand, we propose an anatomical normality modeling method to increase the vision semantic density. 
Our method achieves outstanding zero-shot diagnostic performance on both chest and abdominal CT scenarios and demonstrates excellent transfer learning capabilities in multiple downstream tasks.

\newpage

\noindent\textbf{Acknowledgements}

\noindent{Jianpeng Zhang was supported by the Zhejiang Province Postdoctoral Research Excellence Funding Program (ZJ2024032). This work was also supported by the Zhejiang Provincial “Spearhead \& Pathfinder + X” R\&D Breakthrough Program (2024C03043), Zhejiang Provincial Natural Science Foundation of China (2024-KYI-00I-I05), and National Science Foundation for Distinguished Young Scholars (62225605).}

{
    \small

    \bibliographystyle{ieeenat_fullname}
}
\clearpage
\setcounter{page}{1}
\maketitlesupplementary

\section{More ablation studies}
\subsection{Variety in visual encoder selection}
In the 3D CT VLP task, we discover that the CNN visual encoder outperforms the ViT. Consequently, we explore the impact of various CNN backbones on model performance. As illustrated in Table 6, both ResNet34 and ResNet50 demonstrate improved performance compared to ResNet18. However, considering the balance between computational cost and performance, we decide to utilize ResNet18 as the primary visual encoder in this study.

\subsection{Different initializations for visual encoders}
Aligned with Figure 3, Table 7 provides a numerical comparison of different initialization methods for visual encoder. The table clearly shows that the model initialized with weights derived from our proposed disease-level contrastive learning method achieves the highest AUC, outperforming the other two initialization approaches. These quantitative results further underscore the effectiveness of the proposed visual semantic enhancement.

\subsection{Experiments on local and diffuse diseases}
We assessed the improvement offered by the proposed model over the baseline model in diagnosing both local and diffuse diseases. 
A radiologist categorizes these abnormalities into local and diffuse diseases, as listed in Table~\ref{sup_tab:cls_local_diffuse}.
Detailed performances are presented in Table~\ref{tab:local_diffusion}. As indicated in the table, there is a 4.0\% increase in the AUC for local diseases, which surpasses the 2.8\% improvement seen in diffuse diseases. This suggests that our approach significantly improves the model's ability to diagnose localized diseases.

\section{More implementation details}
For the MedVL-CT69K dataset, we utilize the pre-trained BERT-base~\cite{devlin2018bert} as the text encoder. Our ViSD-boost is trained with the Adam optimizer, where the learning rate increases linearly to 1e-4 in the first epoch and then decreases gradually to 1e-6 via a cosine decay scheduler. The model is trained over four phases for 60, 30, 60, and 30 epochs, utilizing 4 A100 GPUs and a batch size of 48. During training, we dynamically apply RandomCrop and RandomFlip augmentations. For the chest CT-RATE dataset, we employ the same image pre-processing methodology as CT-CLIP~\cite{hamamci2024foundation} to ensure a fair comparison with other methods. We also use the same CXR-Bert as the text encoder~\cite{hamamci2024foundation}. Furthermore, in line with the fVLM~\cite{shui2025large}, we adopt the same anatomy and report parsing methods, facilitating anatomy-wise image-report alignment.

\section{More visualizations of semantic density}
\begin{figure*}[t!]
    \centering
    \includegraphics[width=1.0\linewidth]{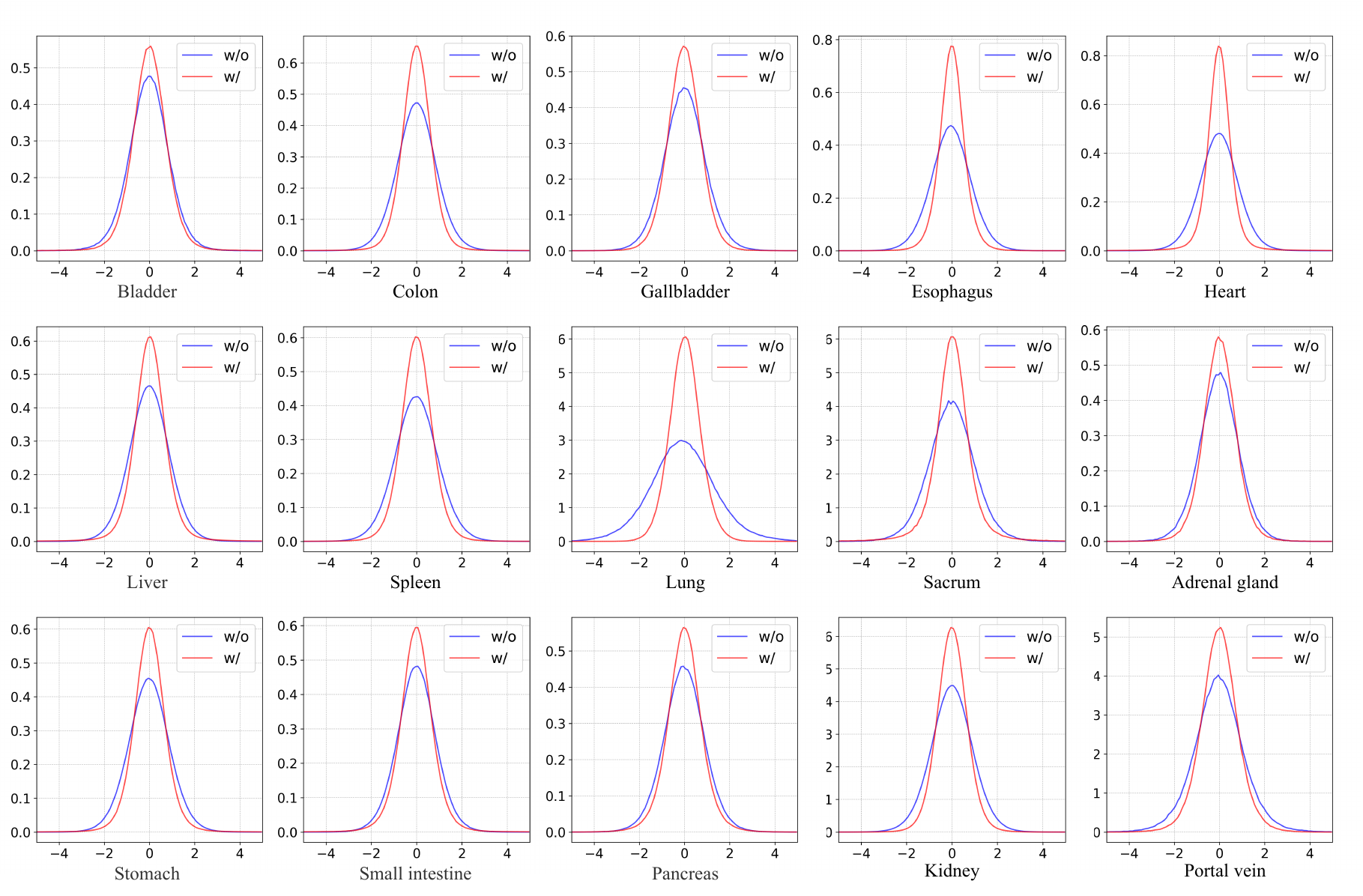}
    \caption{Vision semantic density comparison between models w/ and w/o VSDB.}
    \label{sup_fig:sup_density_all}
\end{figure*}

We present the distributions of visual tokens across additional anatomical structures, as illustrated in Figure~\ref{sup_fig:sup_density_all}. The figure clearly demonstrates that, for all organs, the visual tokens of the model exhibit increased sparsity after the implementation of VSDB, indicating that the model is prioritizing more important features.

\section{Details about zero-shot performance}
\label{sup_sec:detail_performance}
Table~\ref{sup_tab:zero_shot_result} displays the zero-shot performance of the proposed method across 54 abnormalities spanning 15 distinct anatomies.

\begin{table}[]
\centering
\setlength{\tabcolsep}{4.2mm}{
\begin{tabular}{c|cccc}
\toprule
Methods    & SE   & SP   & ACC  & AUC   \\ \midrule
ResNet18   & 73.6 & 75.9 & 74.8 & 78.7 \\
ResNet34   & 74.8 & 76.1 & 75.5 & 78.9      \\
ResNet50   & 76.0 & 75.3 & 75.7 & 79.0      \\  \bottomrule
\end{tabular}}
\caption{Zero-shot performance comparison of different vision encoders on MedVL-CT69K validation set.}
\label{tab:pretrain_contrast}
\end{table}

\begin{table}[]
\centering
\begin{tabular}{c|cccc}
\toprule
Methods    & SE   & SP   & ACC  & AUC   \\ \midrule
Random     & 75.9 & 73.6 & 74.8 & 78.7 \\
Supervised & 76.4 & 75.4 & 75.9 & 79.4      \\
Ours (Disease-level CLP)       & 77.9 & 76.6 & 77.3 & 80.7 \\ \bottomrule
\end{tabular}
\caption{Zero-shot performance comparison of different initialization solutions for vision encoder on MedVL-CT69K validation set. CLP: Contrastive Learning Pre-training.}
\label{tab:pretrain_contrast}
\end{table}

\begin{table}[t]
\begin{tabular}{c|c|cccc|c}
\toprule
Types                    & Methods & SE   & SP  & ACC  & AUC  & $\Delta$               \\ \midrule
\multirow{2}{*}{Local}   & Base    & 72.5 & 70.9 & 71.7 & 76.3 & \multirow{2}{*}{4.0}   \\
                         & Ours    & 75.4 & 74.5 & 75.0 & 80.3 &                      \\ \midrule
\multirow{2}{*}{Diffuse} & Base    & 78.7 & 81.7 & 80.2 & 85.2 & \multirow{2}{*}{2.8} \\
                         & Ours    & 82.1 & 83.1 & 82.6 & 88.0 &                      \\ \bottomrule
\end{tabular}
\caption{Comparison between the base model and our model regarding performance improvements in local and diffuse diseases.}
\label{tab:local_diffusion}
\end{table}

\begin{table*}[t!]
\begin{tabular}{c|c}
\hline
Type      & Diseases  \\ \hline
Local     & \begin{tabular}[c]{@{}c@{}}kidney cyst, kidney stone,   adrenal gland nodule, stomach cancer, gallstone, pancreatic cancer, \\      small intestine diverticulum, small intestine intussusception, colon   cancer, rectal cancer, colon diverticulum, \\      colon appendicolith, liver cyst, liver cancer, liver abscess, liver   metastase, spleen infarction, \\      spleen hemangioma, bladder diverticulum, bladder stone, esophageal varicose   veins, sacrum osteitis\end{tabular}  \\ \hline
Diffuse & \begin{tabular}[c]{@{}c@{}}colon obstruction, colonic gas, colon effusion, colon appendicitis, small intestine obstruction, small intestine gas, \\ small intestine fluid accumulation, cardiomegaly, pericardial effusion, liver glisson’s capsule effusion, liver cirrhosis, \\ intrahepatic bile duct dilatation, fatty liver, bronchiectasis, emphysema, pneumonia, pleural effusion, atelectasis, \\ kidney atrophy, hydronephrosis, adrenal hypertrophy, gastric wall thickening, cholecystitis, pancreatitis, \\ pancreatic duct dilatation, pancreas steatosis, pancreas atrophy, splenomegaly, portal vein hypertension, \\ portal vein thrombosis, esophageal hiatal hernia, gallbladder adenomyomatosis\end{tabular} \\ \hline
\end{tabular}
\caption{Classification of local and diffuse diseases.}
\label{sup_tab:cls_local_diffuse}
\end{table*}

\begin{table*}[t!]

\centering

\resizebox{0.65\textwidth}{!}{

\begin{tabular}{c|c|cccc}

		\toprule

		Anatomy  & Abnormality & SE & SP & ACC & AUC \\

		\hline

		\multirow{2}{*}{Adrenal gland}

		& Adrenal Hypertrophy & 61.5 & 66.3 & 63.9 & 68.0

		 \\

		& Nodule & 65.5 & 63.6 & 64.6 & 68.9

		 \\

		\hline

		\multirow{2}{*}{Bladder} & Diverticulum & 71.4 & 78.8 & 75.1 & 81.6

		 \\

		& Stones & 78.6 & 69.9 & 74.2 & 80.9

		 \\

		\hline

		\multirow{8}{*}{Colon} & Colonic Gas & 74.4 & 81.5 & 78.0 & 85.3

		 \\

		& Effusion & 80.0 & 81.8 & 80.9 & 85.3

		 \\

		& Obstruction & 100 & 95.4 & 97.7 & 99.3

		 \\

		& Diverticulum & 75.0 & 61.1 & 68.0 & 72.3

		 \\

		& Colon Cancer & 77.1 & 68.8 & 72.9 & 80.8

		 \\

		& Rectal Cancer & 80.8 & 88.7 & 84.8 & 92.9

		 \\

		& Appendicitis & 68.4 & 75.3 & 71.9 & 75.8

		 \\

		& Appendicolith & 64.9 & 57.9 & 61.4 & 63.7

		 \\

		\hline

		\multirow{2}{*}{Esophagus} & Hiatal Hernia & 100.0 & 88.3 & 94.2 & 96.9

		 \\

		&  Varicose Veins & 98.7 & 97.1 & 97.9 & 99.6

		 \\

		\hline

		\multirow{3}{*}{Gallbladder} & Cholecystitis & 67.1 & 69.7 & 68.4 & 74.4

		 \\

		& Gallstone & 68.2 & 79.4 & 73.8 & 80.4

		 \\

		& Adenomyomatosis & 61.7 & 60.0 & 60.9 & 63.0

		 \\

		\hline

		\multirow{2}{*}{Heart} & Cardiomegaly & 90.0 & 91.4 & 90.7 & 97.0 \\

		& Pericardial Effusion & 79.2 & 74.1 & 76.6 & 84.1

		 \\

		\hline

		\multirow{4}{*}{Kidney} & Atrophy & 78.4 & 89.6 & 84.0 & 89.5

		 \\

		& Cyst & 62.7 & 62.2 & 62.5 & 67.5

		 \\

		& Hydronephrosis & 85.1 & 84.4 & 84.7 & 89.9

		 \\

		& Renal Calculus & 63.5 & 61.9 & 62.7 & 67.1

		 \\

		\hline

		\multirow{8}{*}{Liver}& Fatty Liver & 84.0 & 78.4 & 81.2 & 90.4

		 \\

		& Glisson’s Capsule Effusion & 89.7 & 84.8 & 87.2 & 93.8

		\\

		& Metastase  & 73.8 & 82.8 & 78.3 & 86.4

		 \\

		& Intrahepatic Bile Duct Dilatation  &74.6 & 73.1 & 73.9 & 80.4

		 \\

		& Cancer  & 86.9 & 89.3 & 88.1 & 93.8

		 \\

		& Cyst & 61.0 & 54.3 & 57.6 & 61.0

		 \\

		& Abscess & 66.7 & 92.7 & 79.7 & 85.6

		 \\

		& Cirrhosis  & 90.4 & 87.2 & 88.8 & 96.0

		 \\

		\hline

		\multirow{5}{*}{Lung} & Atelectasis & 95.6 & 95.9 & 95.8 & 99.0

		 \\

		& Bronchiectasis & 94.4 & 85.6 & 90.0 & 96.2

		 \\

		& Emphysema  & 80.0 & 79.7 & 79.8 & 79.0

		 \\

		& Pneumonia  & 81.1 & 82.0 & 81.6 & 88.9

		 \\

		& Pleural Effusion & 95.7 & 95.8 & 95.7 & 98.2

		 \\

		\hline

		\multirow{5}{*}{Pancreas} & Pancreatic Cancer & 93.1 & 82.6 & 87.8 & 94.7

		 \\

		& Atrophy & 83.8 & 86.1 & 84.9 & 91.1

		 \\

		& Pancreatitis & 85.7 & 93.6 & 89.6 & 95.2

		 \\

		& Pancreatic Duct Dilatation & 58.5 & 84.7 & 71.6 & 77.8

		 \\

		& Steatosis & 82.2 & 78.8 & 80.5 & 85.7

		 \\

		\hline

		\multirow{2}{*}{Portal vein} & Hypertension & 94.4 & 90.8 & 92.6 & 97.9

		 \\

		& Thrombosis & 90.9 & 94.5 & 92.7 & 96.7

		 \\

		\hline

		\multirow{5}{*}{Small Intestine} & Gas Accumulation & 81.9 & 83.2 & 82.6 & 89.3

		 \\

		& Fluid Accumulation & 80.3 & 82.3 & 81.3 & 87.9

		 \\

		& Obstruction & 85.2 & 86.9 & 86.1 & 92.9

		 \\

		& Diverticulum & 84.1 & 77.1 & 80.6 & 88.2

		 \\

		& Intussusception & 88.9 & 66.1 & 77.5 & 83.4

		 \\

		\hline

		\multirow{3}{*}{Spleen} & Hemangioma & 68.1 & 65.9 & 67.0 & 69.4

		 \\

		&  Infarction & 81.8 & 81.9 & 81.9 & 87.5

		 \\

		&  Splenomegaly & 84.7 & 83.6 & 84.1 & 91.8

		 \\

		\hline

		\multirow{2}{*}{Stomach} &  Gastric Wall Thickening & 67.5 & 74.4 & 70.9 & 77.7

		 \\

		& Gastric Cancer & 78.6 & 80.3 & 79.5 & 84.5

		 \\

		\hline

		Sacrum & Osteiti & 70.6 & 75.7 & 73.2 & 77.5 \\

		\bottomrule

		\multicolumn{2}{c|}{Average}  & 79.4 & 79.6 & 79.5 & 84.9 \\

		\bottomrule

\end{tabular}}

\caption{Detailed zero-shot performance of our method on each abnormality.}

\label{sup_tab:zero_shot_result}

\end{table*}

\end{document}